\documentclass[12pt,preprint]{aastex}

\shorttitle{A Planet Around GJ 176}
\shortauthors{Endl et al.}

\begin{document}

\title{An $m \sin i = 24$ Earth Mass Planetary Companion To The Nearby M Dwarf GJ~176.
\footnote{Based on observations obtained with the Hobby-Eberly Telescope, which is a joint project of the 
University of Texas at Austin, the Pennsylvania State University, Stanford University, 
Ludwig-Maximilians-Universit\"at M\"unchen, and Georg-August-Universit\"at G\"ottingen.}}

\author{Michael Endl and William D. Cochran}
\affil{McDonald Observatory, The University of Texas at Austin,  
    Austin, TX 78712}
\email{mike@astro.as.utexas.edu, wdc@astro.as.utexas.edu}

\author{Robert A. Wittenmyer}
\affil{Astronomy Department, The University of Texas at Austin,
    Austin, TX 78712}
\email{robw@astro.as.utexas.edu}

\and

\author{Alan P. Boss}
\affil{Department of Terrestrial Magnetism, Carnegie Institution of Washington,
Washington, D.C. 20015-1305}
\email{boss@dtm.ciw.edu}

\begin{abstract}

We report the detection of a planetary companion with a minimum mass of 
$m \sin i = 0.0771~{\rm M}_{\rm Jup} = 24.5~{\rm M}_{\oplus}$ to the nearby ($d=9.4$~pc) M2.5V star GJ~176.
The star was observed as part of our M dwarf planet search at the Hobby-Eberly Telescope (HET).  
The detection is based on 5 years of high-precision differential radial velocity (RV) measurements 
using the High-Resolution-Spectrograph (HRS). The orbital period of the planet is $10.24$~d. GJ~176 thus 
joins the small (but increasing) sample of M dwarfs hosting short-periodic planets with minimum masses in 
the Neptune-mass range. Low mass planets could be relatively common around M~dwarfs and the current 
detections might represent the tip of a rocky planet population.   

\end{abstract}


\section{Introduction}

Over the past few years our preliminary knowledge on the frequency of extrasolar
planets in the low mass part of the HR-diagram has increased significantly. An ever increasing number
of M dwarfs is being monitored by various Doppler searches with high radial velocity (RV) precision. 
This has led to several discoveries of M dwarf planets that cover an enormous and surprising range in mass, 
almost comparable to the mass range of the planets in our Solar System. 
Jovian planets were found around GJ~876 (Delfosse et al.~1998, Marcy et al.~1998, 2001), GJ 849 (Butler et 
al.~2006) and GJ 317 
(Johnson et al.~2007), and Neptune-mass planets around GJ~436 (Butler et al.~2004), GJ~581 (Bonfils et al.~2005b) 
and GJ~674 (Bonfils et al.~2007). The low primary masses of M dwarfs combined with state-of-the-art RV 
precision even allowed the detection of additional planets with minimum masses below $10~{\rm M}_{\oplus}$ 
(so-called ``Super-Earths'') in the GJ~876 (Rivera et al.~2005) and GJ~581 (Udry et al.~2007) systems. 
  
In this paper we report the detection of a new low mass planet in a $10.24$~d orbit around the M dwarf 
GJ~176. This is the fourth planet in the class of Neptune-mass companions orbiting M~dwarfs. 
GJ~176 is already the third (previously unknown) planet host in our 
HET M dwarf planet search (Endl et al. 2003, 2006) sample.

\section{Stellar parameters of GJ~176}

GJ~176 (HD 285968, HIP 21932, LHS 196) is a $V=9.97$ M2.5Ve star at a distance of $9.4$ parsec, according to 
the Hipparcos parallax of $106 \pm 2.5$ mas (Perryman et al.~1997). The star has a $B-V$ of $1.51$ and 
an absolute $V$ magnitude of $10.08$. The 2MASS magnitudes for GJ~176 are $J=6.462$, $H=5.824$ and 
$K=5.607$ mag (Cutri et al.~2003). 
Using the V-band and K-band mass-luminosity relationships of Delfosse et al.~(2000), we estimate a 
mass of $0.48$ and $0.50~{\rm M}_{\odot}$, respectively. We adopt the mean value of $0.49\pm0.014~{\rm 
M}_{\odot}$ as the mass of the star.

A photometric study by Weis~(1994) didn't find significant variability 
for GJ~176. The V-band scatter measured by Weis~(1994) is 0.006 mags, equal to the measurement uncertainties.   
The ROSAT All-Sky-Survey catalog of nearby stars (H\"unsch et al.~1999) reports a moderate coronal X-ray 
emission level of $3 \times 10^{27}\,{\rm erg\,s}^{-1}$. 
GJ~176 is thus a moderately active star, possibly exhibiting starspots and flares, 
a quite typical behavior for M dwarfs. Rauscher \& Marcy~(2006) measured the equivalent widths of the Ca II
H and K lines with $0.97\pm0.10$~\AA~and $0.69\pm0.08$~\AA, respectively. Based on the V-band 
mass-metallicity relationship of Bonfils et al.~(2005a) we estimate an [Fe/H]$=-0.1\pm0.2$ for GJ~176, so 
roughly solar metallicity.

\section{Observations and RV results}

We observed GJ~176 as part of our on-going Doppler search for planets around M dwarfs (Endl et al. 2003, 
2006) using the HET (Ramsey et al.~1998) and its HRS spectrograph (Tull~1998). We started observations
of GJ~176 in 2003 October 15 and collected a total of 28 spectra over 5 years. All observations were
performed using our standard planet search setup and data reduction pipeline decribed in detail in Cochran et 
al.~(2004). We use the common I$_2$ cell technique to obtain high precision differential RV measurements 
(e.g. Butler et al.~1995, Endl et al.~2000). 

\begin{figure}
\includegraphics[angle=270,scale=.60]{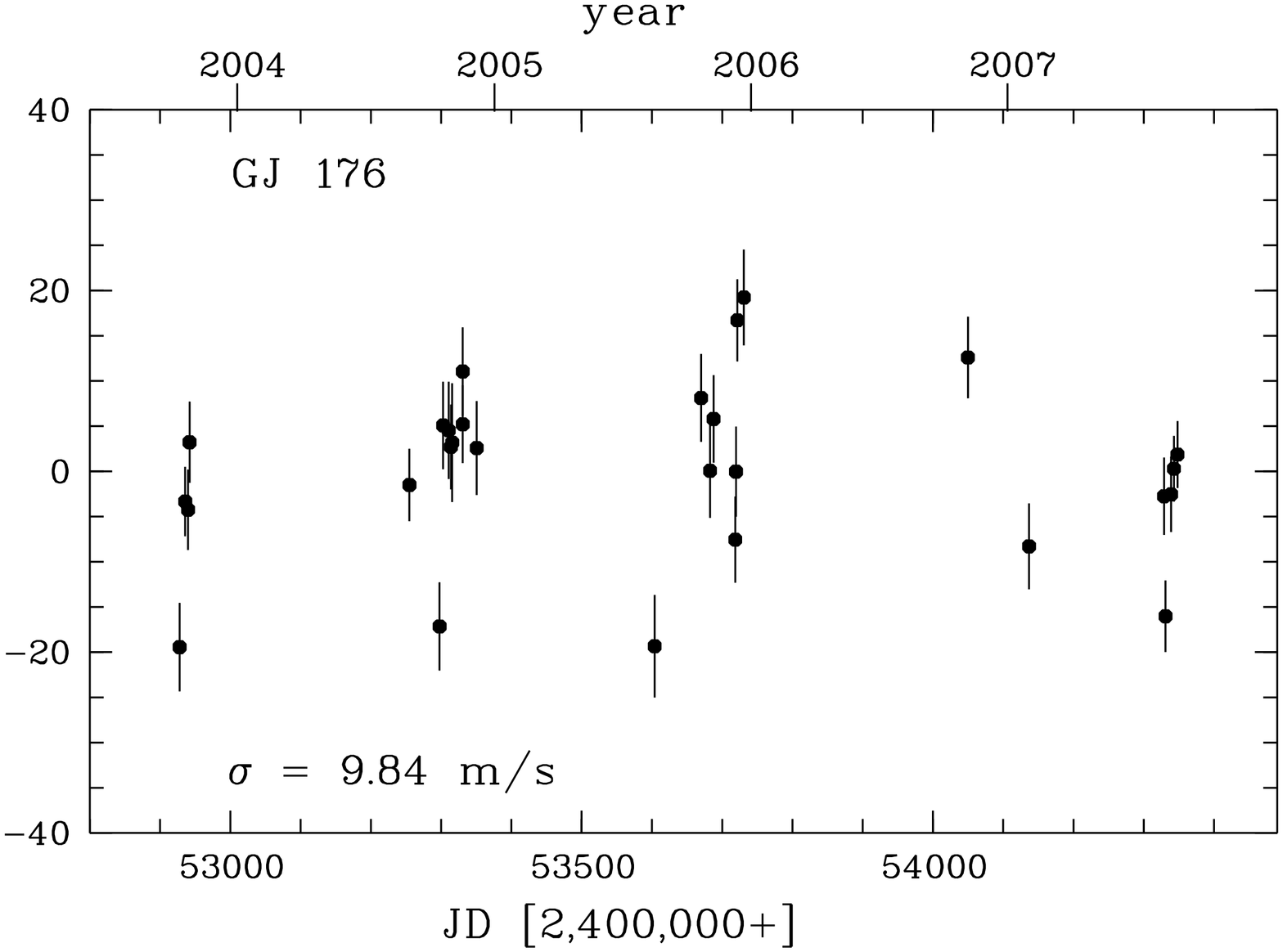}
\caption{5 years of HET/HRS RV measurements of GJ~176. The data have a scatter of $9.84~{\rm m \, s}^{-1}$
and an average measurement uncertainty of $4.69~{\rm m \, s}^{-1}$. 
\label{rvs}}
\end{figure}

Fig.~\ref{rvs} shows the time series of our HET/HRS RV measurements with
the small secular acceleration of the RV of $0.36~{\rm m\,s}^{-1}{\rm yr}^{-1}$, as computed
from the Hipparcos parallax and proper motion information, already subtracted. 
The data have an overall rms-scatter of $9.84~{\rm m\,s}^{-1}$ and average internal errors of 
$4.69\pm0.63~{\rm m\,s}^{-1}$. The total scatter is more than twice the measurement uncertainties,
indicative of intrinsic RV variability of this target. The HET RV data are listed in Table~\ref{data}.

\section{Period search and orbital solution}

\begin{figure}
\includegraphics[angle=270,scale=.60]{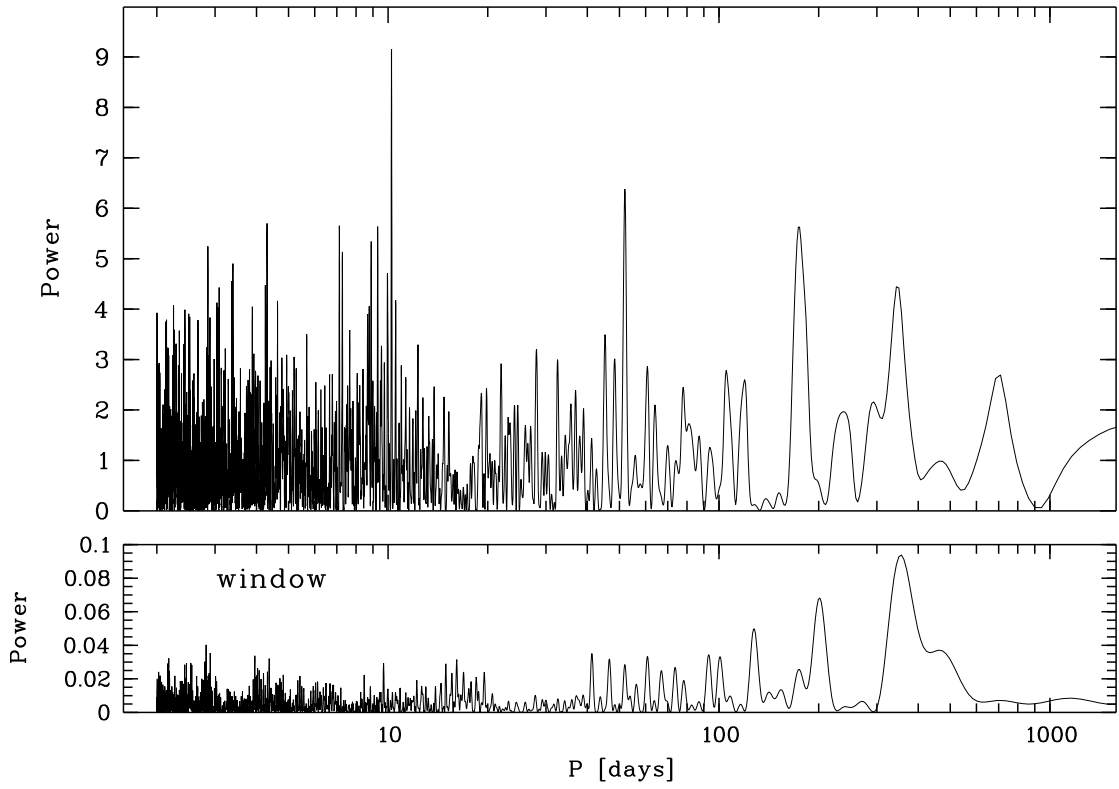}
\caption{Lomb-Scargle periodogram of the RV data of GJ~176.
The highest peak is at $10.24$ days and has a false-alarm-probability
of $0.0004$. The lower panel displays the window function of our observations. \label{lomb}}
\end{figure}

We searched the RV data of GJ~176 for any significant periodicities using the classic
Lomb-Scargle periodogram (Lomb~1976 ; Scargle~1982). Fig.~\ref{lomb} displays the resulting power spectrum.
A strong peak is visible at a period of $10.24$~d. We estimate the false-alarm-probability (FAP) of this
signal with the bootstrap randomization scheme (e.g. K\"urster et al.~1997). After 100,000 bootstrap 
re-shuffling runs we find that the FAP of the $10.24$~d signal is only 0.0004. 

As the next step we use Gaussfit (Jeffereys et al.~1988) to find a Keplerian orbital solution 
to our RV data. A circular orbit fit yields a $\chi^{2}$ of $35.2$ ($\chi^{2}_{\rm red}=1.47$) and
a residual RV scatter of $5.57~{\rm m\,s}^{-1}$. A slightly better fit is obtained with an eccentric
orbit with $e=0.23\pm0.13$: $\chi^{2}$ of $32.86$ ($\chi^{2}_{\rm red}=1.49$) and residual rms of $5.32~{\rm 
m\,s}^{-1}$.
The large uncertainity in $e$ and the fact
that a circular orbit yields a lower $\chi^{2}_{\rm red}$ urges us to remain cautious concerning the reality 
of the non-circular orbit. However, future observations will allow us to determine whether the orbit is 
indeed eccentric. A moderately eccentric orbit could be an indication for additional planets around this 
star and thus warrants intensive follow-up monitoring.

\begin{figure}
\includegraphics[angle=270,scale=.60]{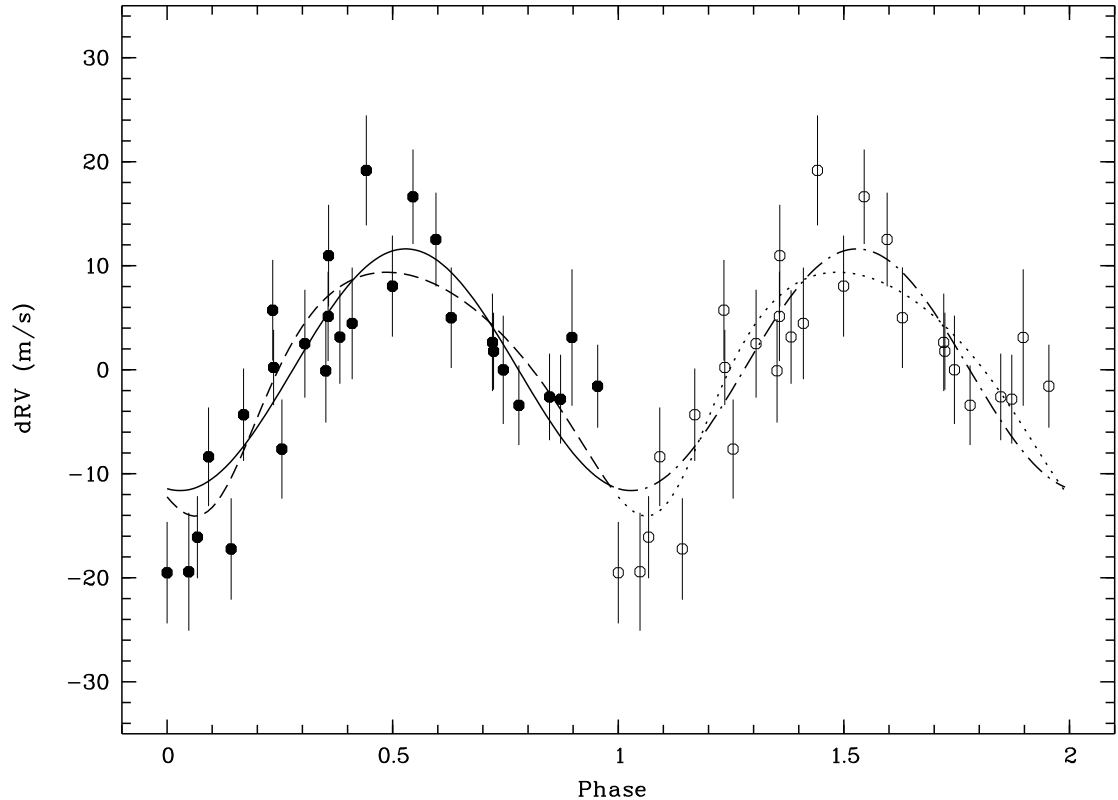}
\caption{The best-fit orbital solutions are shown as solid line (circular) and dashed
line (eccentric) along with the HET/HRS RV measurements
phased to the orbital period of $10.24$~d (the data are plotted twice for a second cycle). 
The semi-amplitude $K$ of $11.6~{\rm m\,s}^{-1}$ corresponds to
a minimum mass of $24.5~{\rm M}_{\oplus}$ for the companion. All parameters of the orbital solution
are summarized in Table~\ref{tab:planet2}. The residual scatter around the fit is $5.57~{\rm m \, s}^{-1}$ 
(circular) and $5.32~{\rm m \, s}^{-1}$ (eccentric orbit).
\label{phase}}
\end{figure}

Fig.~\ref{phase} shows the RV measurements and both orbital solutions (circular and eccentric) phased 
to the best-fit period. The circular orbital solution yields a minimum mass for the companion 
of $0.077\pm0.012~{\rm M}_{\rm Jup} = 24.5\pm3.9~{\rm M}_{\oplus}$, while an eccentric orbit would
lower the minimum mass slightly to $0.076\pm0.010~{\rm M}_{\rm Jup} = 24.1\pm3.1~{\rm M}_{\oplus}$.
The orbital parameters are summarized in Table~\ref{tab:planet2}.

\subsection{Stellar activity versus planet hypothesis}

The case of GJ~674 has demonstrated clearly that star spots can introduce low level signals in high precision
RV data of M dwarfs (Bonfils et al.~2007). Can the GJ~176 signal also be caused by rotational modulation due to star spots
and not a planet? This scenario is very unlikely because a rotation period of $10.2$~d would mean that GJ~176
would rotate more than 3 times faster than GJ~674, and this should lead to a higher activity level than 
GJ~674. This is not the case, as GJ~176 has a slightly lower coronal X-ray emission than GJ~674 (H\"unsch et 
al.~1999). As mention before, Weis~(1994) didn't detect photometric variability in GJ~176 and we also
did not find any significant variability or periodicity in the
Hipparcos photometry for this star (the highest peak
at $3.7$~d has a FAP of $1.5\%$.)
 
Moreover, the vast majority of the data of Bonfils et al.~(2007) was collected over a relatively short period of time ($\approx 
200$~d), where rotational modulation can mimic a Keplerian signal of a planet because the active regions on the star  
remain constant for this period of time. However, over a larger amount of time, active regions will 
reconfigure and emerge at different stellar longitudes causing a phase as well as amplitude change in the RV 
signal. The GJ~176 signal remains stable in phase and amplitude over 5 years, which makes the planet 
hypothesis much more plausible as the origin of the RV signal.

\section{Discussion}

Remarkably, GJ~176 is already the third M dwarf with a Neptune-mass companion that was included in our 
HET sample of $60$ M dwarfs. GJ~436 and GJ~581 were part of our M dwarf sample before the planets around them
were announced. The frequency of short-periodic Neptune-mass planets around M dwarfs in our HET
sample is hence $5\%$, which is higher than the frequency of hot Jupiters around FGK-type stars 
of $\approx 1.2\%$ (Marcy et al.~2005). Neptunes and Super-Earths could be relatively common around low
mass stars. Of course, with M dwarfs we are still limited by small number statistics, as compared to the
few thousands of FGK-type stars already observed by various Doppler surveys around the world.

We are currently increasing the size of the HET M dwarf sample, and combined with the
results of other programs observing M dwarfs, we should be able to derive important constraints 
for planet formation models. The present
information about Neptune-mass planets on short-period orbits suggests
that they may be the tip of the terrestrial planet distribution, for
several reasons. First, a number of hot Neptunes have 2 or 3 sibling gas
giant planets orbiting at much greater distances: $\rho$~1~Cnc, $\mu$~Ara, and GJ~876 
all have such systems. This planetary system architecture is the same
as our own solar system, with inner terrestrial planets and outer gas
giant planets, and suggests that the hot Neptunes in these systems formed
inside their gas giants, making them likely to be rocky planets.
Theoretical models of the collisional accumulation of terrestrial planets
predict that rocky planets as massive as about $21~{\rm M}_{\oplus}$ would
result from a protoplanetary disk with a surface density of gas and solids
high enough to form a gas giant planet by core accretion in a few million
years (Wetherill~1996; Inaba et al.~2003). Formation of the Neptunes
interior to their gas giants seems to be a much more likely scenario than
formation as ice giants on orbits outside the orbits of the gas giants,
followed by migration somehow past the gas giants to their current
short-period orbits. While GJ~876 is the only M dwarf of these three
stars, one might expect a common explanation for its hot Neptune as a
rocky world, as well as those of GJ~436 and GJ~581. This interpretation
bodes well for the eventual detection of Earth-mass planets on habitable
orbits around low mass stars. For M~dwarfs with masses of 
$<0.2~{\rm M}_{\odot}$ we already have the sensitivity to detect 
$m \sin i \approx 1~{\rm M}_{\oplus}$ planets in the habitable zone 
(Endl \& K\"urster~2007, in prep.).

While M dwarfs appear to have hot Neptunes at a higher rate than G dwarfs,
they appear to have short-period gas giant planets at a somewhat
smaller rate than G dwarfs (Endl et al.~2006 ; Johnson et al.~2007).  
Similarly, microlensing
surveys have detected planetary companions at asteroidal distances to
M dwarf stars at frequencies that suggest that planets at such locations
are more likely to be Neptune-mass than Jupiter-mass. These results
may simply be a result of M dwarfs have lower mass protoplanetary disks
than G dwarfs, with the consequent result that the planets that form tend
to have a mass distribution that is shifted downward. However, this
apparent preference for cold Super-Earths over Jupiters can also be
explained at present by both the core accretion and disk instability
formation mechanism for gas giant planets. In the case of core accretion,
the lengthened orbital periods around M dwarfs could lead to most cores
growing too slowly to accrete significant gaseous envelopes, making failed
cores the usual outcome, rather than gas giants  (Laughlin et al.~2004).
Disk instability explains the appearance of both gas giants and Super-Earths
at asteroidal distances around M dwarfs by appealing to rapid formation
of gaseous protoplanets by disk instability, followed by conversion to
ice giants (cold Super-Earths) by photoevaporation at asteroidal distances
in region of high mass star formation (Boss~2006). Because most M dwarfs
form in regions of high mass formation, they should be orbited at asteroidal
distance primarily by Super-Earths. M~dwarfs that form in region of
low mass star will have gas giant planets at those distances, if disk
instability is operative (Boss 2006). It remains for these theoretical
speculations to be tested by completing the extrasolar planetary census.
      
\acknowledgments
This material is based on work supported by the National Aeronautics and
Space Administration under Grants NNG04G141G, NNG05G107G issued through
the Terrestrial Planet Finder Foundation Science program and Grant
NNX07AL70G issued through the Origins of Solar Systems Program.
The Hobby-Eberly Telescope (HET) is a joint project of the University of
Texas at Austin, the Pennsylvania State University, Stanford University,
Ludwig-Maximilians-Universit\"{a}t M\"{u}nchen,
and Georg-August-Universit\"{a}t G\"{o}ttingen.
The HET is named in honor of its principal benefactors,
William P. Hobby and Robert E. Eberly. We would like to thank the
McDonald Observatory TAC for generous allocation of observing time.

%{\it Facilities:} \facility{Nickel}, \facility{HST (STIS)}, \facility{CXO (ASIS)}.

\clearpage

\clearpage

\clearpage

%\clearpage

\begin{deluxetable}{rrrrrr}
\tabletypesize{\tiny}
\tablecaption{
Differential radial velocities for GJ~176 from the HET/HRS}
\tablewidth{0pt}
\tablehead{
\colhead{JD [2,400,000+]} & \colhead{dRV [${\rm m\,s}^{-1}$]} &
\colhead{$\sigma$ [${\rm m\,s}^{-1}$]} &
\colhead{JD [2,400,000+]} & \colhead{dRV [${\rm m\,s}^{-1}$]} &
\colhead{$\sigma$ [${\rm m\,s}^{-1}$]}
\label{data} }
\startdata
52927.82553 &     -19.88   &    4.88 & 53669.99475 &  7.67  &     4.86 \\
52935.80776 &      -3.78   &     3.83 & 53682.74654 & -0.38 &   5.20 \\
52939.79789   &    -4.69   &     4.44& 53687.75180 & 5.35 &  4.84\\
52941.98273    &    2.78    &    4.49& 53718.67107 &      -8.00   &     4.76\\
53254.93830   &    -1.95  &      3.99& 53719.66762 &      -0.47   &     4.98\\
53297.80620  &    -17.60   &     4.88& 53721.64443  &     16.26   &     4.54\\
53302.79940  &      4.63   &     4.84& 53730.82258  &     18.79   &     5.29\\
53310.78943  &      4.09   &     5.37& 54049.74686  &     12.15  &      4.51\\
53313.97346  &      2.26   &     4.69& 54136.72087  &     -8.74   &     4.74\\
53315.77643  &      2.73   &     6.56& 54328.97056  &     -3.20   &     4.27\\
53330.71871  &      4.76   &     4.30& 54330.96817  &    -16.47   &     3.95\\
53330.72892  &     10.60   &     4.89& 54338.95808  &     -2.98   &     4.16\\
53350.66330  &      2.13   &     5.19& 54342.93427  &     -0.15   &     3.62\\
53603.95259 &      -19.79 &       5.67 & 54347.92539  &      1.41  &      3.70\\
\enddata
\end{deluxetable}

\clearpage

\begin{deluxetable}{lr@{$\pm$}lr@{$\pm$}l}
\tablecolumns{5}
\tablewidth{0pt}
\tablecaption{Parameters of GJ~176~b for circular (left column) and eccentric orbit (right column)
\label{tab:planet2}}
\tablehead{
\colhead{Parameter} &
\multicolumn{2}{c}{Value}
& \multicolumn{2}{c}{Value}
}
\startdata
P & 10.2369 & 0.0039 days & 10.2366 & 0.0038 days\\
T & 2454550.6672 & 0.39 BJD& 2455037.7979 & 1.1 BJD\\
K     & 11.62 & 1.61 m\,s$^{-1}$& 11.72 & 1.62 m\,s$^{-1}$\\
e     & 0.0 & 0.0 (fixed)& 0.232 & 0.127\\
$\omega$ & 0.0 & 0.0 degrees (fixed)& 210.4 & 32.5 degrees\\
\hline
$M \sin i $ & 0.0771 & 0.0122 $M_{\rm Jup}$& 0.0757 & 0.0096 $M_{\rm Jup}$\\
$M \sin i $ & 24.5 & 3.9 $M_{\oplus}$& 24.1 & 3.1 $M_{\oplus}$\\
a & 0.0727 &  0.0007 AU& 0.0727 &  0.0007 AU\\
RMS &  \multicolumn{2}{c}{5.57 m\,s$^{-1}$}&  \multicolumn{2}{c}{5.32 m\,s$^{-1}$}\\
%\hline
\enddata
\end{deluxetable}

\end{document}